\documentclass{article}
\pdfoutput=1

\usepackage{arxiv}
\usepackage[utf8]{inputenc} 
\usepackage[T1]{fontenc}    
\usepackage{hyperref}       
\usepackage{url}            
\usepackage{booktabs}       
\usepackage{amsfonts}       
\usepackage{nicefrac}       
\usepackage{microtype}      
\usepackage{lipsum}		    

\usepackage[american]{babel}
\usepackage{csquotes}
\usepackage{amsmath}
\usepackage{todonotes}
\usepackage{tikz}
\usepackage{float}
\usepackage{lineno}
\usepackage{multirow}
\usepackage{adjustbox}
\usepackage{amsthm}
\usepackage{amssymb}
\usepackage{graphicx}
\usepackage[figuresright]{rotating}
\usepackage{comment}
\usepackage{setspace}

\usepackage{relsize}
\usepackage[flushleft]{threeparttable}
\usepackage[style=apa,sortcites=true,sorting=nyt,backend=biber]{biblatex}
\DeclareLanguageMapping{american}{american-apa}
\title{Assessing quality of selection procedures:\\Lower bound of false positive rate as a function of inter-rater reliability}
\renewcommand{\shorttitle}{Reliability and binary classification framework}
\author{František Bartoš$^{1,2}$, Patrícia Martinková$^{1,3}$ \\
\small $^{1}$ Institute of Computer Science of the Czech Academy of Sciences, Prague, Czech Republic\\
\small $^{2}$ University of Amsterdam, Amsterdam, Netherlands \\
\small $^{3}$ Faculty of Education, Charles University, Prague, Czech Republic\\}

\begin{document}

\maketitle

\thispagestyle{empty}

\begin{abstract}
Inter-rater reliability (IRR) is one of the commonly used tools for assessing the quality of ratings from multiple raters. However, applicant selection procedures based on ratings from multiple raters usually result in a binary outcome; the applicant is either selected or not. This final outcome is not considered in IRR, which instead focuses on the ratings of the individual subjects or objects. We outline the connection between the ratings' measurement model (used for IRR) and a binary classification framework. We develop a simple way of approximating the probability of correctly selecting the best applicants which allows us to compute error probabilities of the selection procedure (i.e., false positive and false negative rate) 
or their lower bounds.
We draw connections between the inter-rater reliability and the binary classification metrics, showing that binary classification metrics depend solely on the IRR coefficient and proportion of selected applicants. We assess the performance of the approximation in a simulation study and apply it in an example comparing the reliability of multiple grant peer review selection procedures. We also discuss possible other uses of the explored connections in other contexts, such as educational testing, psychological assessment, and health-related measurement and implement the computations in \texttt{IRR2FPR} \texttt{R} package.
\end{abstract}

\newpage

\clearpage
\pagenumbering{arabic}

\section{Introduction}

High-quality applicant, grant, or manuscript (jointly referred to as applicant thereafter) selection procedures are necessary for fair and efficient allocation of resources. Selection procedures usually aim to assess applicants' quality, which can be thought of as a latent variable, based on observable (and potentially imperfect) indicators of their quality \parencite{bartholomew2011latent}. 

Selection procedures coarsen ratings into a binary outcome: an applicant may be rated on a ten-point scale, but in the end is either hired or refused, a grant proposal is either funded or dismissed, and an article is either accepted or rejected \parencite[e.g.,][]{cicchetti1991reliability}. However, the most commonly used metric for evaluating the reliability of selection procedures---the inter-rater reliability (IRR)---focuses on the ratings rather than the outcomes of selection procedures.\footnote{This is a different goal than measuring \emph{agreement} of multiple raters performing binary decisions using e.g., Cohen's Kappa \parencite{cohen1960coefficient}, that, again, measures agreement and not whether the best applicants were selected.} 

This is, of course, not a novel observation, and the disconnection between evaluating raters' reliability and whether the best applicants were selected was noted earlier \parencite[e.g.,][]{kraemer1991we, nelson1991process, mayo2006peering}. In cases where applicant selection is based on a fixed threshold (i.e., pass/fail tests), the expected classification accuracy can be estimated by methods outlined by \textcite{rudner2000computing, rudner2005expected} and \textcite{guo2006expected} \parencite[also see][]{lee2010classification, livingston1995estimating, hanson1990investigation}. We extend the aforementioned approach to settings where a proportion of the best candidates is selected and show that, under the assumption of a normally distributed latent variable, the expected classification accuracy can be directly obtained from IRR and the proportion of selected candidates. Subsequently, the selection procedures can be characterized as binary classification and evaluated via well-known and interpretable metrics such as sensitivity or false positive/negative rates. Furthermore, the binary classification framework allows researchers and stakeholders to evaluate and improve selection procedures while incorporating the costs of incorrect decisions, increasing the number of raters, or modifying the rating procedure. In fact, connecting reliability to binary classification recalls classical models that evaluate selection procedures based on validity \parencite[e.g.,][]{taylor1939relationship, cronbach1957psychological}.

While our approach is related to univariate classification under measurement error, this use case differs. Compared to typical classification tasks, which aim to separate subjects into different categories \parencite{duda2006pattern}, we assume the existence of a single continuous latent trait (or a composite score based on a multidimensional assessment) measured by the ratings. Then, we aim to select the best applicants defined by the latent trait and evaluate the (miss)classification probabilities due to the measurement error contained in the observed ratings. Ideally, the validity of the observed indicators (and their combination into the overall assessment) would be evaluated directly, thus answering the question of how well the results of selection procedures predict applicants' success. However, such a ``gold standard'' measure of success needed for directly assessing the validity of selection procedure is often not available (\cite{lauer2015reviewing, moher2016increasing, superchi2019tools}, also see \cite{grant2022refinement} for alternatives); either because the success measure is located too far in the future, or it is difficult to agree on the success measure itself (see \cite{lauer2015predicting, li2015big} for suggestion to use bibliometric measures in grant reviews, and \cite{fang2016nih, lindner2015examining} for a subsequent critique). Consequently, we are often left to assess the reliability of the selection procedures, as reliability limits the usefulness even of completely valid indicators \parencite{thurstone1931reliability}.  Therefore, our approach results in the lower bound on the corresponding error probabilities as it does not account for additional miss-classifications due to (a lack of) validity. 

The paper proceeds as follows: First, we define a minimal measurement model for ratings in the selection procedures and characterize the selection procedures in terms of their outcome using the binary classification. Then we propose a quantile approximation connecting the measurement model with the binary classification and show the relationship between IRR and the binary classification metrics. Second, we evaluate the quality of the quantile approximation in a simulation study by comparing the empirical true positive rate to the estimated true positive rate. Third, we demonstrate the outlined methodology on an example of multiple grant peer review, where we compare multiple selection procedures in terms of their false positive rate, false negative rate, and $\text{F}_1$ score and then show how the false positive rate changes with increasing the number of raters and IRR. Finally, we close with a discussion of limitations and extensions.

\section{Methods}
\label{sec:methods}

\subsection{Measurement Model}
\label{sec:measurement_mmodel}

For ease of exposition, we start with a simple selection procedure that evaluates $N$ applicants with the goal of estimating their latent abilities $\gamma_i$, which would traditionally be analyzed with a one-way ANOVA model (i.e., assuming either the lack of rater overlap or lack of rater effect on the ratings). More complicated scenarios, for example, rater effects, when the same set of raters performs the ratings, would be accounted for via a two-way ANOVA or more complex models (see \cite{brennan2001, martinkova2023computational}). In the selection process, each applicant is rated $J$ times, resulting in a vector of observed scores $y_{ij}$ for $i = 1, \dots N$ and $j = 1, \dots, J$. The resulting measurement model can be written as 
\begin{equation}
    \label{eq:model}
    y_{ij} = \gamma_{i} + \epsilon_{ij},
\end{equation}
with the standard assumptions that (a) the measurement error $\epsilon_{ij}$ of each applicant is independently normally distributed with zero mean and variance $\sigma^2_{\epsilon}$ and (b) the latent abilities $\gamma_i$ are normally distributed with mean $\mu$ and variance $\sigma^2_{\gamma}$ \parencite{searle2006variance, deleeuw2008handbook}. Precision of the estimates depends on the appropriateness of the specified model. In the ``Simulation'' section, we explore the effect of deviations from the specified model on the results.

The quality of the selection procedure is usually measured as inter-rater reliability (sometimes also referred to as single-rater IRR), which is the intra-class correlation coefficient \parencite[ICC(1,1)][]{mcgraw1996forming, shrout1979intraclass} 
\begin{equation}
    \label{eq:IRR1}
    \nonumber
    \text{IRR}_1 = \frac{\sigma_\gamma^2}{\sigma_\gamma^2 + \sigma_\epsilon^2}.
\end{equation}
When the final decisions are based on the average of $J$ raters, the multiple-rater IRR can be calculated using the Spearman-Brown formula as 
\begin{equation}
    \label{eq:IRRJ}
    \nonumber
    \text{IRR}_J = \frac{\sigma_\gamma^2}{\sigma_\gamma^2 + \frac{\sigma_\epsilon^2}{J}} = \frac{J  \, \text{IRR}_1}{J  \, \text{IRR}_1 + 1 - \text{IRR}_1}.
\end{equation}

\subsection{Binary Classification}
\label{sec:binary_classification}

The endpoint of the selection procedure is, however, not estimating the latent abilities $\gamma_i$ per se but selecting $k$ applicants with the highest latent ability $\gamma_i$. Consequently, we can divide the applicants into two groups: high-ability applicants $A$ that consist of $k$ applicants with the highest latent ability $\gamma_i$ and ``not high-ability applicants'' $\neg A$ that consist of the remaining $N-k$ applicants. This is similar to \textcite{taylor1939relationship} who differentiate between ``satisfactory'' and ``unsatisfactory'' applicants. The difference between the described and the \textcite{taylor1939relationship} framing is that we are interested in selecting the ``best'' rather than ``satisfactory'' applicants---a much more stringent requirement, which simplifies the problem and allows us to evaluate the procedure without the knowledge of the ``satisfactory'' cutoff. Denoting the selected applicants by $S$ and the remaining applicants by $\neg S$, we can characterize the selection procedure by probabilities of the different types of (miss)classification (Table~\ref{tab:classification}).

\begin{table}[h]
    \centering
    \begin{tabular}{r|cc|c}
                                   &  Selected ($S$)     & Not Selected ($\neg S$)      &     \\
    \hline
    High-Ability Applicants ($A$)          &  $P(S \cap A)$      & $P(\neg S \cap A)$           & $P(A)$     \\
    Not High-Ability Applicants ($\neg A$) &  $P(S \cap \neg A)$ & $P(\neg S \cap \neg A)$      & $P(\neg A)$  \\
    \hline
                                   &  $P(S)$             & $P(\neg S)$                  & $1$     \\
    \end{tabular}
    \caption{Overview of the selection procedure as a binary classification.}
    \label{tab:classification}
\end{table}

Since the proportion (i.e., unconditional probability) of selected applicants corresponds, by definition, to the proportion  (i.e., unconditional probability) of high-ability applicants, $P(S) = P(A) = \nicefrac{k}{N}$, the whole classification process is determined by the probability of correctly selecting the high-ability applicants $P(S \cap A)$, i.e., by the probability of true positives classifications (Table~\ref{tab:classification2}).

\begin{table}[h]
    \centering
    \begin{tabular}{r|cc|c}
                                   &  Selected ($S$)                  & Not Selected ($\neg S$)              &                       \\
    \hline
    High-Ability Applicants ($A$)          &  $P(S \cap A)$                   & $\nicefrac{k}{N} - P(S \cap A)$      & $\nicefrac{k}{N}$     \\
    Not High-Ability Applicants ($\neg A$) &  $\nicefrac{k}{N} - P(S \cap A)$ & $P(S \cap A) + \nicefrac{(N-2k)}{N}$ & $1 - \nicefrac{k}{N}$ \\
    \hline
                                   &  $\nicefrac{k}{N}$               & $1 - \nicefrac{k}{N}$                & $1$                   \\

    \end{tabular}
    \caption{Simplification of the binary classification in selection procedures.}
    \label{tab:classification2}
\end{table}

Subsequently, if we knew the probability of true positive classification, we could use standard metrics for evaluating binary classification \parencite[e.g.,][pp. 14-33]{pepe2003statistical}. Since the off-diagonal probabilities of the classification outcomes are equal (Table~\ref{tab:classification2}), many commonly specified metrics become equal or simplify to a constant. For example, true positive rate (TPR, aka sensitivity) and positive predictive value (PPV, aka precision) are equal, 
\begin{equation}
    \label{eq:TPR}
    \text{TPR} = \text{PPV} = \frac{P(S \cap A)}{\nicefrac{k}{N}},
\end{equation}
and true negative rate (TNR, aka specificity) is equal to $\nicefrac{1}{2}$. We can also compute the false positive rate (FPR, corresponding to type I error rate), 
\begin{equation}
    \label{eq:FPR}
    \text{FPR} = \frac{\nicefrac{k}{N} - P(S \cap A)}{\nicefrac{k}{N}},
\end{equation}
or false negative rate (FNR, corresponding to type II error rate),
\begin{equation}
    \label{eq:FNR}
    \text{FNR} = \frac{\nicefrac{k}{N} - P(S \cap A)}{\nicefrac{(N - k)}{N}}.
\end{equation}
Some more complex functions of classification probabilities, e.g., an $F_1$ score, also become identical to the true positive rate,
\begin{equation}
\nonumber    F_1 = \frac{\text{TP}}{\text{TP} + \nicefrac{1}{2} (\text{FP} + \text{FN})} = \frac{P(S \cap A)}{P(S \cap A) + \nicefrac{1}{2} (\nicefrac{k}{N} - P(S \cap A) + \nicefrac{k}{N} - P(S \cap A))} = \frac{P(S \cap A)}{\nicefrac{k}{N}}.
\end{equation}
The classification probabilities can be used with a utility function specifying the cost of each classification type \parencite[e.g.,][]{metz1978basic}.

\subsection{Quantile Approximation}
\label{sec:quantile_approximation}

The probability of true positive classification is, unfortunately, not directly estimable from the observed scores $y_{ij}$, unless we know the true latent ability $\gamma_i$ of each applicant. To deal with this hindrance, we propose a quantile approximation that allows us to estimate the true positive classification probability indirectly. Specifically, we leverage distributional assumptions of the measurement model (Equation~\ref{eq:model}), our ability to estimate the populational parameters of the measurement model, and one peculiarity of the selection procedure: the fact that we set the unconditional probability of the applicant with high ability, $P(A)$ and the unconditional probability of the applicant selected, $P(S)$, when designing the selection procedure.

In selection procedures, the high-ability applicants are defined by their ranking on the latent ability $\gamma_i$. However, we only measure the observed scores $y_{ij}$ when performing the selection. According to the measurement model, the observed scores $y_{ij}$ are an imprecise reflection of the latent abilities $\gamma_i$, which is disturbed by some measurement error with variance $\sigma^2_\epsilon$. The $J$ observed scores of each applicant are usually aggregated into a mean score $\Bar{y}_i$ with the resulting measurement error $\sigma_\epsilon/\sqrt{J}$. Importantly, the latent abilities $\gamma_i$ and the random variable of the aggregated scores $\Bar{y}_i$ are jointly distributed with bivariate normal density:
\begin{equation}
    \label{eq:joint_distribution}
    \begin{bmatrix} \gamma_i \\ \Bar{y}_i \end{bmatrix}
    \sim N\left(
    \begin{bmatrix} \mu \\ \mu \end{bmatrix},  \begin{bmatrix} \sigma_\gamma^2 & \sigma_\gamma^2 \\ \sigma_\gamma^2 & \sigma_\gamma^2 + \sigma_\epsilon^2 / J \end{bmatrix}
    \right),
\end{equation}
\parencite[e.g.,][pp. 258-259]{searle2006variance}.

Since the goal of selection procedures is selecting the high-ability $k$ applicants and since the desirability of applicants is defined by their latent ability $\gamma_i$, the high-ability applicants are defined as those with the highest $k$ latent ability scores $\gamma_i$. Then, we can think about a cut-score on the latent ability $\text{g}_\text{c}$ that separates the high-ability applicants from the remaining applicants (with the high-ability applicants satisfying the condition $\gamma_i > \text{g}_\text{c}$). Similarly, since the selection procedure is performed with the observed aggregated scores $\Bar{y}_i$ and since we select the high-ability $k$ applicants according to their observed aggregated scores $\Bar{y}_i$, the selected applicants are defined as those with the highest $k$ observed scores $\Bar{y}_i$. Then, again, we can think about a cut-score on the observed aggregated scores $\Bar{\text{y}}_\text{c}$ that separates the high-ability applicants from the remaining applicants (with the high-ability applicants satisfying the condition $\Bar{y}_i > \Bar{\text{y}}_\text{c}$).

In any particular data set, the cut-scores $\text{g}_\text{c}$ and $\Bar{\text{y}}_\text{c}$ are dependent on the actual latent abilities $\gamma_i$ and observed aggregated scores $\Bar{y}_i$ of applicants participating in the given selection procedure. However, under the assumption of the measurement model, we can approximate the cut-scores with marginal quantile functions of the joint distribution (Equation~\ref{eq:joint_distribution}), yielding the quantile approximated cut-scores 
\begin{align}
    \label{eq:cut_scores}
    \Tilde{\text{g}}_\text{c} &= \mu + \sigma_\gamma \Phi^{-1}\left(\frac{N-k}{N}\right), \\
    \nonumber
    \Tilde{\Bar{\text{y}}}_\text{c} &= \mu + \sqrt{\sigma_\gamma^2 + \sigma_\epsilon^2 / J} \, \Phi^{-1}\left(\frac{N-k}{N}\right),
\end{align}
with $\Phi$ corresponding to a cumulative distribution function of a standard normal distribution.

Subsequently, the true positives classification probability in Table~\ref{tab:classification2}, defining the binary classification metrics, can be approximated using the latent abilities $\gamma_i$, its imperfect measurement via the random variable of aggregated observed scores $\Bar{y}_i$, and the quantile approximated cut-scores $\Tilde{\text{g}}_\text{c}$ and $\Tilde{\Bar{\text{y}}}_\text{c}$,
\begin{equation}
    \nonumber
    P(S \cap A) \approx P(\gamma > \Tilde{\text{g}}_\text{c}, \Bar{y} > \Tilde{\Bar{\text{y}}}_\text{c}).
\end{equation}
In other words, the probability of an applicant being amongst both the high-ability and selected applicants can be approximated by the probability of both having latent ability higher than the approximated cut-score on the latent ability and having the observed aggregated score higher than the approximated aggregated cut-score.

Finally, the true positives classification probability can be approximated solely with the populational parameters and integration over the joint bivariate normal density
\begin{equation}
    \label{eq:correct_classification}
    P(\gamma > \Tilde{\text{g}}_\text{c}, \Bar{y} > \Tilde{\Bar{\text{y}}}_\text{c}) = \mathop{\mathlarger{{\int}}}_{\Tilde{\text{g}}_\text{c}}^\infty \mathop{\mathlarger{{\int}}}_{\Tilde{\Bar{\text{y}}}_\text{c}}^\infty
    N\left(
    \begin{bmatrix} \gamma \\ \Bar{y} \end{bmatrix}
    \mathop{\mathlarger{\mathlarger{\mathlarger{\mid}}}}
    \begin{bmatrix} \mu \\ \mu \end{bmatrix},  \begin{bmatrix} \sigma_\gamma^2 & \sigma_\gamma^2 \\ \sigma_\gamma^2 & \sigma_\gamma^2 + \sigma_\epsilon^2 / J \end{bmatrix}
    \right) \, d \Bar{y}  \, d \gamma.
\end{equation}

With the true positive classification probability at hand, we can assess the selection procedure in the binary classification framework and compute its error rates and other binary classification metrics.

The true positive classification probability approximation is correct to the extent to which the observed scores $y_{ij}$ are a valid measure of the underlying ability $\gamma_i$. Often, the observed ratings are not only affected by lacking reliability but also by lacking validity. In such a case, the true positive classification probability is overestimated (i.e., additional errors occur due to the lack of validity), and the computed miss-classification metrics (FPR and FNR) become an upper bound on the true miss-classification metrics.

This approach is similar to the tetrachoric correlation coefficient \parencite[e.g.,][]{pearson1900mathematical, bonett2005inferential}, which aims to estimate the correlation between latent continuous variables based on observed dichotomized variables instead of assessing the classification of latent dichotomous variables based on observed continuous variables.

\subsection{Relationship between IRR and Binary Classification}
\label{sec:IRR_and_binary_classification}

We can further utilize the quantile approximation to draw connections between IRR and the true positives classification probability of the binary classification.

We subtract the grand mean $\mu$ from both the latent abilities $\gamma_i$ and the observed aggregated scores $\Bar{y}_i$ and standardize the random variables by the total variance of the observed aggregated scores $\sigma_\gamma^2 + \sigma_\epsilon^2 / J$. This transforms the latent abilities into $\theta_i = (\gamma_i - \mu) / \sqrt{\sigma_\gamma^2 + \sigma_\epsilon^2 / J}$ and the observed aggregated scores into $\Bar{z}_i = (\Bar{y}_i - \mu) / \sqrt{\sigma_\gamma^2 + \sigma_\epsilon^2 / J}$. Subsequently, the bivariate normal density of the transformed abilities and the random variable of transformed observed aggregated scores are simplified to
\begin{equation}
\label{eq:joint_distribution2}
    \nonumber
    \begin{bmatrix} \theta_i \\ \Bar{z}_i \end{bmatrix}
    \sim N\left(
    \begin{bmatrix} 0 \\ 0 \end{bmatrix},  \begin{bmatrix} \text{IRR}_J & \text{IRR}_J \\ \text{IRR}_J & 1 \end{bmatrix}
    \right),
\end{equation}
and yields the quantile approximated cut-scores of the transformed abilities and observed aggregated measures
\begin{align}
    \label{eq:cut_scores_z}
    \nonumber
    \Tilde{\text{t}}_\text{c} &= \text{IRR}_J \Phi^{-1}\left(\frac{N-k}{N}\right), \\
    \nonumber
    \Tilde{\Bar{\text{z}}}_\text{c} &=  \Phi^{-1}\left(\frac{N-k}{N}\right),
\end{align}
which also highlights that the two cut-scores are equal under perfect inter-rater reliability, $\text{IRR}_J = 1$.

The true positives classification probability can be directly estimated with only $\text{IRR}_J$ and the proportion of selected applicants $k$,
\begin{equation}
    \label{eq:correct_classification_IRR}
    \nonumber
    P(\theta > \Tilde{\text{t}}_\text{c}, \Bar{z} > \Tilde{\Bar{\text{z}}}_\text{c}) = \mathop{\mathlarger{{\int}}}_{\Tilde{\text{t}}_\text{c}}^\infty \mathop{\mathlarger{{\int}}}_{\Tilde{\Bar{\text{z}}}_\text{c}}^\infty
    N\left(
    \begin{bmatrix} \theta \\ \Bar{z} \end{bmatrix}
    \mathop{\mathlarger{\mathlarger{\mathlarger{\mid}}}}
    \begin{bmatrix} 0 \\ 0 \end{bmatrix},  \begin{bmatrix} \text{IRR}_J & \text{IRR}_J \\ \text{IRR}_J & 1 \end{bmatrix}
    \right) \, d \Bar{z}  \, d \theta.
\end{equation}
This allows us to retrospectively evaluate selection procedures in the binary classification framework without access to the primary data.

\section{Simulation}
\label{sec:simulation}

We conducted a simulation study to assess the performance of the quantile approximation. The simulation considers four data-generating scenarios to assess how the quantile approximation is affected by the violation of its assumptions; (1) `normal`, where the data-generating model corresponds to the assumptions of the quantile approximation, (2) `skew`, where the true abilities are positively skewed, (3) `bias`, where a proportion of applicants is rated in a reverse matter, and (4) `dependent`, where ratings across participants are dependent on the reviewers. The simulation compared the empirical true positive classification probabilities $P(S \cap A)$ based on the simulated abilities $\gamma_i$ vs. the aggregated ability estimates $\Bar{y}_i$ (i.e., the grand truth under complete knowledge) with the true positives classification probability obtained by quantile approximation $P(\gamma > \Tilde{\text{g}}_\text{c}, \Bar{y} > \Tilde{\Bar{\text{y}}}_\text{c})$ based on $\Hat{\sigma}_\gamma^2$ and $\Hat{\sigma}_\epsilon^2$ estimates from a linear mixed model (Equation~\ref{eq:correct_classification}). We focus on the true positive classification probabilities rather than the error rates, or other classification metrics, as the error rates are derived quantities of the true positive classification probability.

In the `normal` scenario, the data were simulated from the measurement model defined by Equation~\eqref{eq:model} with settings inspired by \textcite{martinkova2018disparities}. We manipulated the single inter-rater reliability (intra-class correlation coefficient) $\text{IRR}_1 = \{0.15, 0.30, 0.45\}$ (fixing the overall variance $\sigma^2$ to 1),\footnote{Whereas the measurement model is defined in terms of the grand mean $\mu$ and variances $\sigma_\gamma^2$ and $\sigma_\epsilon^2$, Equation~\eqref{eq:correct_classification_IRR} shows that $\text{IRR}_J$ is the only relevant quantity for the quantile approximation} number of applicants, $N = \{100, 300, 1000\}$, and the number of ratings, $J = \{3, 5, 10\}$. The range of $\text{IRR}_1$ values corresponds to values found in grand peer review evaluations (see Table~\ref{tab:grant_reviews} in the Example section) and the higher number of raters create settings where the total inter-rater reliability of the rating reaches $\text{IRR}_J = 0.89$ (setting with $\text{IRR}_1 = 0.45$ and $J = 10$).

The remaining scenarios were generated by the following modifications of the  `normal` scenario; the `skewed` scenario simulated the latent abilities $\gamma_i$ from Gamma(3, 3) distribution which generated a notable right skew, the `biased` scenario reversed ratings for 10\% of the applicants, i.e., breaking the relationship between mean ratings and the latent abilities $\gamma_i$, and the `dependent` scenario generated ratings from non-exchangeable raters where between-rater variance constituted 0.10 of the total variance.

We replicated each simulation condition 1000 times and estimated the variance parameters $\sigma_\gamma^2$ and $\sigma_\epsilon^2$ using linear models implemented in the \texttt{lme4} \texttt{R} package \parencite[Version 1.1.35.1,][]{lme4}. In the remainder of the section, we only discuss results based on $N = 100$ applicants, but see (Appendix A) for similar results with $N = 300$ and $N = 1000$.

\begin{figure}[h!]
    \centering
    \includegraphics[width=\textwidth]{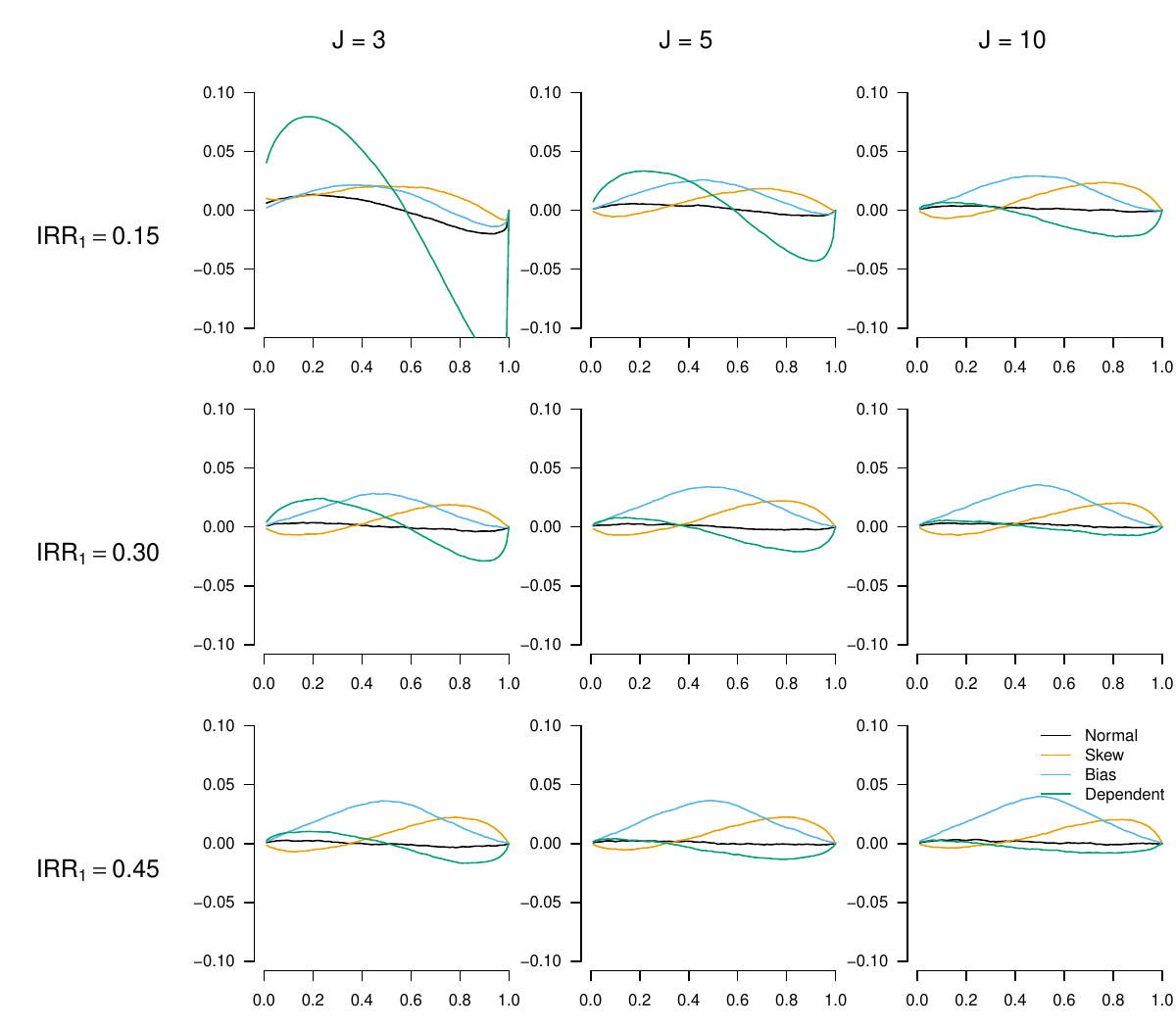}
    \caption{Bias of quantile approximated true positive classification probabilities for 100 applicants across all possible proportions of selected applicants ($x$-axis) with different data generating scenarios encoded with color. Results are presented for different inter-rater reliability coefficients (rows) and the number of ratings (columns).}
    \label{fig:sim_N=100}
\end{figure}

Figure~\ref{fig:sim_N=100} visualizes the bias of quantile approximated true positive classification probabilities across all possible proportions of selected applicants ($\nicefrac{k}{N}$, $x$-axis) with different data generating scenarios encoded with color. The results are presented for different inter-rater reliability coefficients ($\text{IRR}_1$, rows), the number of ratings ($J$, columns), and 100 applicants. The `normal` (black) scenario showed little to no bias across all conditions. The `skew` (orange) scenario resulted in a negative bias of the true positive classification probabilities for a small proportion of applicants and a positive bias for a large proportion of applicants for all but the lowest number of ratings and the lowest single inter-rater reliability (due to estimation issues of the between-applicant variance). The `bias` (blue) scenario showed steadily increasing positive bias for up to half of the selected applicants, with a steadily decreasing bias after that. The `dependent` (green) scenario resulted in more distinct and opposite bias than the `skew` scenario, i.e., positive of the true positive classification probabilities for a small proportion of applicants and negative bias for a large proportion of applicants which was extremely pronounced in the lowest number of ratings and the lowest single inter-rater reliability. Across all but the `bias` scenario, the magnitude of bias decreased with the increasing number of ratings and the single inter-rater reliability. Increased sample size also reduced the degree of bias although to a smaller degree.

\begin{figure}[h!]
    \centering
    \includegraphics[width=\textwidth]{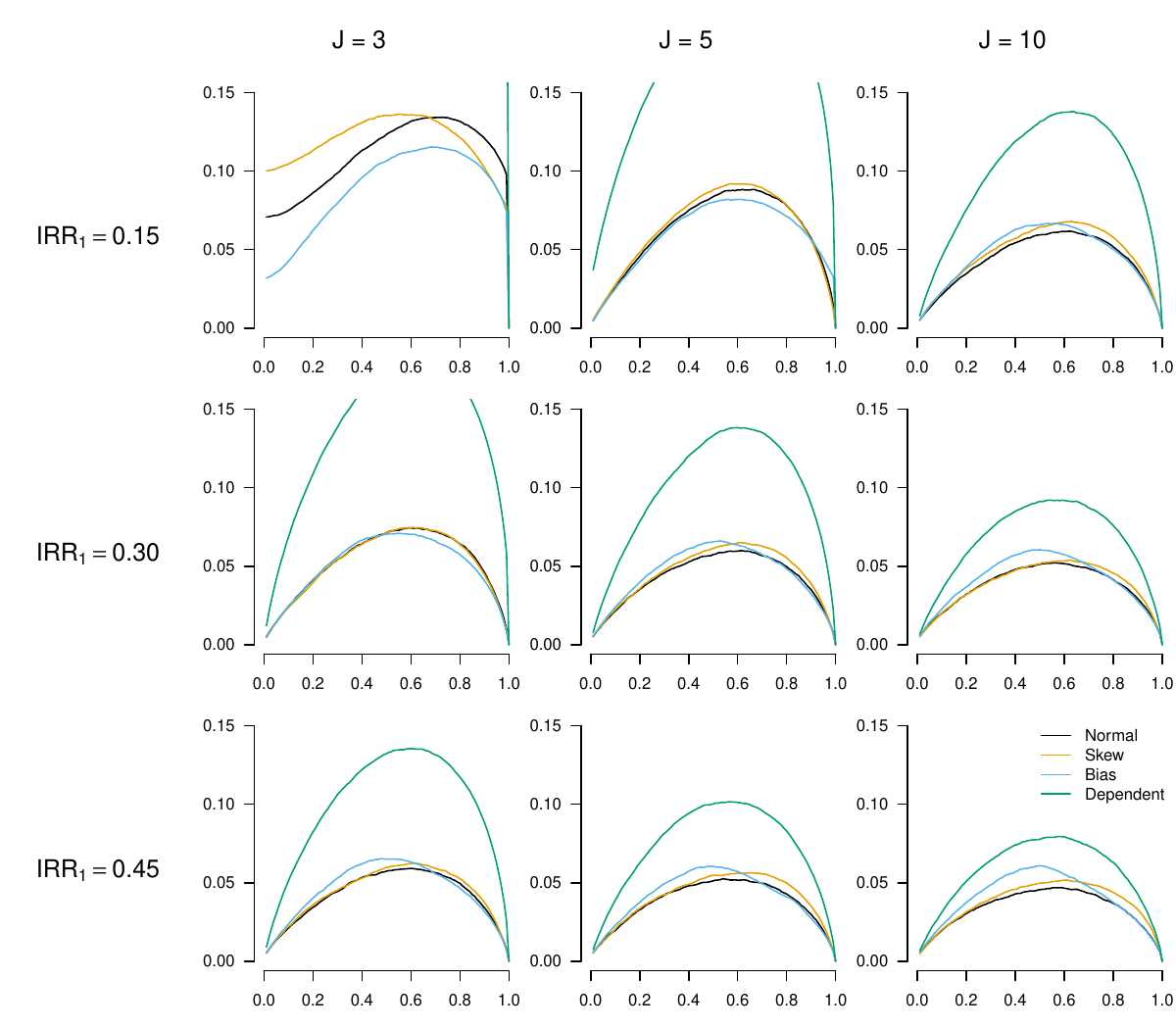}
    \caption{Root mean square error of quantile approximated true positive classification probabilities for 100 applicants across all possible proportions of selected applicants ($x$-axis)with different data generating scenarios encoded with color. Results are presented for different inter-rater reliability coefficients (rows) and the number of ratings (columns).}
    \label{fig:RMSE_N=100}
\end{figure}

Importantly, the quantile approximation estimates the expected classification probabilities for a selection procedure with given populational characteristics. In other words, the quantile approximation might provide noisy estimates of the true values for any particular selection procedure, especially when the classification probabilities depend on a small number of selected applicants. Figure~\ref{fig:RMSE_N=100} visualizes the root mean square error (RMSE) of quantile approximated true positive classification probabilities across all possible proportions of selected applicants ($\nicefrac{k}{N}$, $x$-axis) with different data generating scenarios encoded with color. The results are presented for different inter-rater reliability coefficients ($\text{IRR}_1$, rows), number of ratings ($J$, columns), and 100 applicants. The $x$-axis corresponds to the proportion of selected applicants ($\nicefrac{k}{N}$). Note that the true positive classification probability is bounded at the endpoints---it is zero if no candidate is selected, and it is one if all candidates are selected. As such, the RMSE of the true positive classification probability is necessarily higher above the middle proportion of selected applicants since it can attain the widest range of values. We find that the RMSE is comparable in the `normal`, `skew`, and `bias` scenarios. With 1000 observations, we can notice an increased RMSE for the `skew` scenario in high proportions of selected applicants, which is a result of the biased estimates. The `dependent` scenario produces a much larger degree of RMSE, especially in conditions with a low number of ratings and single inter-rater reliability. The RMSE of the approximation, again, improves with the increasing number of ratings and single inter-rater reliability as well as with the increasing number of applicants.

\subsection{Confidence and Prediction Intervals}

\begin{figure}[h!]
    \centering
    \includegraphics[width=0.5\textwidth]{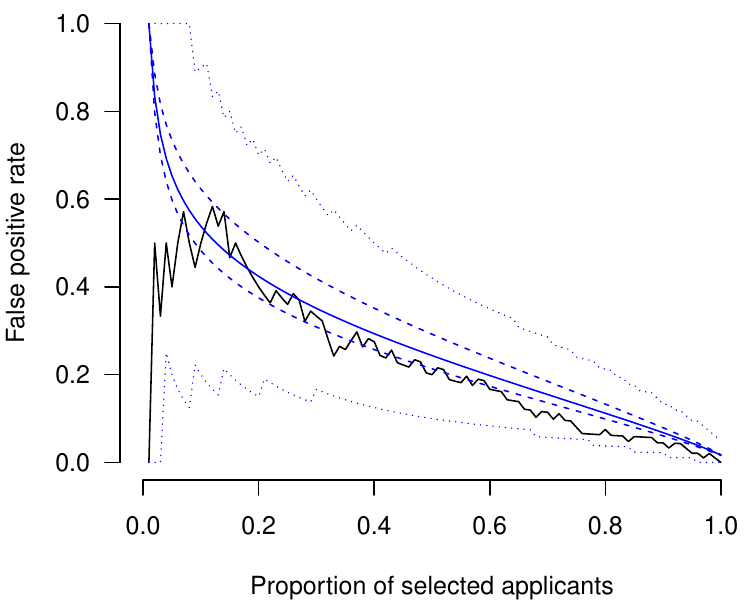}
    \caption{Example of quantile approximated estimate of the false positive rate (full blue line), the 95\% confidence interval (blue dashed line), 95\% prediction interval (blue dotted line), and the empirical false positive rate (full black line) from a random simulation with 5 ratings, $\text{IRR}_1 = 0.30$, and 100 applicants.}
    \label{fig:sim_single}
\end{figure}

Dependency on any particular selection procedure can be further illustrated by visualizing the empirical vs. quantile approximated results for a single trial. For example, consider the false positive rate, the probability an applicant is selected while they are not part of the high-ability group (a transformation of the previously summarized true positive classification probability, Equation~\ref{eq:FPR}). Figure~\ref{fig:sim_single} compares the quantile approximated false positive rate (full blue line) to a random simulation's empirical false positive rate (full black line) with $J = 5$, $\text{IRR}_1 = 0.30$, and $N = 100$.

The empirical false positive rate wildly oscillates between 0 and 1, especially when considering the selection of only a small number of applicants ($x$-axis). These wild oscillations, accompanied by the extremely wide 95\% prediction interval (blue dotted line; based on quantile function of binomial distribution), are an inherent property of summarizing the proportion of successes of a binomial outcome with a small number of trials. In contrast, 95\% confidence intervals of the false positive rate (blue dashed lines), which quantifies the uncertainty of the false positive rate \emph{estimate} (based on non-parametric bootstrap) is wider in the lower proportion of the selected candidates and shorter around the endpoints. This results from the false positive rate estimate being bounded at the endpoints, with higher uncertainty in the lower proportion of selected applications---as it is a transformation of the true positive classification probability.

\section{Example: Grant Peer Reviews}
\label{sec:example}

We illustrate the methodology by estimating binary classification metrics for several grant peer reviews. See Table~\ref{tab:grant_reviews} for characteristics of the grant peer reviews as summarized in Table 1 of \textcite{erosheva2021jrss} and extended by the results reported therein.\footnote{We focus only on studies that estimated $\text{IRR}$ using all applicants to prevent bias from the restricted range \parencite{erosheva2021jrss}.} It is important to note that the reported values are averages, often aggregated across disciplines and years (sometimes accompanied by a range of values). The results should, therefore, be taken as an illustration showcasing the possible interpretation and inferences rather than an evaluation of the funding agencies' grant review process. 

\begin{table}[h]
    \centering
    \begin{tabular}{rcccc}
    Study & Proposals & $\text{IRR}_1$ & $J$ & Selected \\
    \hline
    \textcite{cicchetti1991reliability}    & 150 NSF \& COSPUP (1985) & $0.18 - 0.37$ & $4.24$ \& $4.04$ & $51\%^1$      \\
    \textcite{mutz2012heterogeneity}       & 8329 FWF (1999-04)       & $0.26$        & $2.82$           & $32\%-53\%$   \\
    \textcite{carpenter2015retrospective}  & 260 AIBS (2009-11)       & $0.14 - 0.41$ & $2$              & $5\%-11\%$    \\
    \textcite{erosheva2021jrss}            & 72 AIBS (2014-17)        & $0.37$        & $3$              & $34\%-38\%$   \\
    \textcite{erosheva2021jrss}            & 2076 NIH (2014-16)       & $0.34$        & $2.79$           & $18\%^2$        \\
    \end{tabular}
    \caption{Overview of the grant peer reviews}
    \begin{tablenotes}
        \small
        \item $^1$ Based on Table 46 from \textcite[p. 140]{cole1978peer}.
        \item $^2$ Based on \textcite{erosheva2020nih}.
        \item COSPUP = Committee on Science and Public Policy of the National Academy of Sciences, NSF = National Science Foundation, FWF = Austrian Science Fund, AIBS = American Institute of Biological Sciences, NIH = National Institutes of Health.
        \item $J$ corresponds to the mean number of raters in \textcite{cicchetti1991reliability}, \textcite{mutz2012heterogeneity}, and \textcite{erosheva2021jrss} as the number of raters per proposal was unequal.
    \end{tablenotes}
    \label{tab:grant_reviews}
\end{table}

\begin{figure}[h!]
    \centering
    \includegraphics[width=1\textwidth]{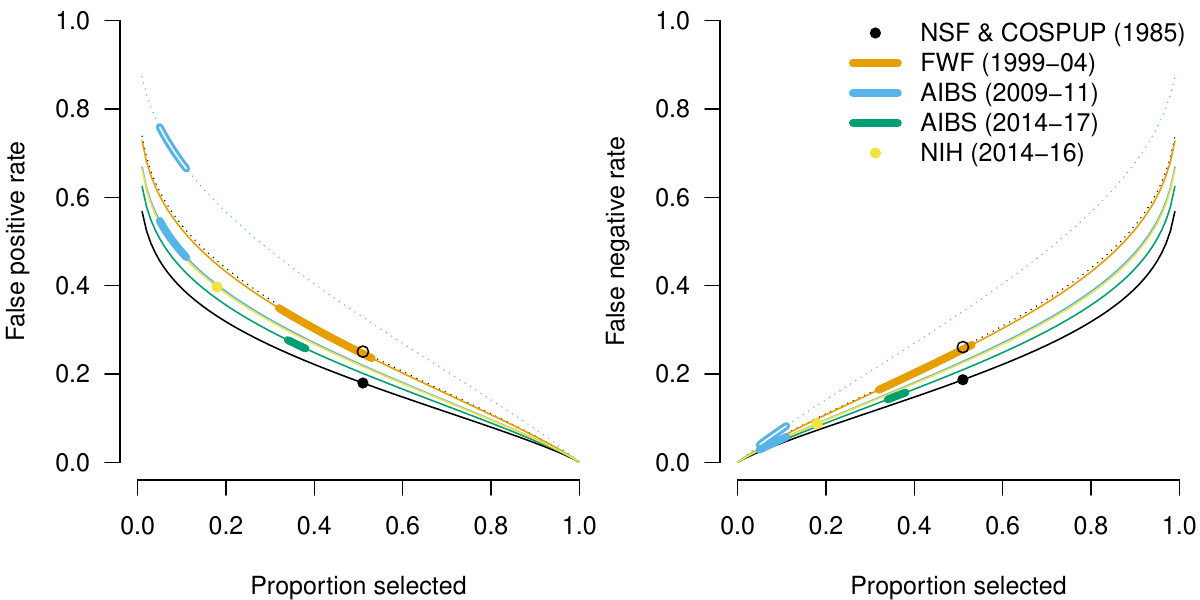}
    \caption{Estimates of the false positive and false negative rate for grant peer review procedures based on quantile approximation for published estimates of IRR and number of raters $J$ of individual studies (different colors). Thin full lines visualize the estimated false positive rate across the range of possible proportions of selected applicants, and thick full lines (and points) highlight the ranges (and values) of the actual proportion of selected applicants. For NSF \& COSPUP and AIBS, the lower $\text{IRR}_J$ estimate, resulting in a higher false positive rate, is shown as dotted lines / empty circles, whereas the higher $\text{IRR}_J$ estimate is shown as full lines / full circles.}
    \label{fig:example_errors}
\end{figure}

Figure~\ref{fig:example_errors} visualizes the false positive rate (left) and false negative rate (right) estimates for each grant peer-review procedure (different colors) based on quantile approximation. Thin full lines visualize the computed false positive rate across the range of possible proportions of selected applicants, and thick full lines (and points) highlight the ranges (and values) of the actual proportion of selected applicants. The estimated false positive rate ranges from $0.75$ for the lowest proportion of selected applicants ($5\%$), lowest inter-rater reliability ($0.14$), and two rates in the AIBS 2009-11 grant proposals data to $18\%$ for the second highest proportion of selected applicants ($51\%$), high inter-rater reliability ($0.37$) and more than four raters in the NSF \& COSPUP (1985) grant proposals data. However, the false negative rate shows the opposite pattern; the lowest false negative rates of $0.029$ and $0.040$ are for the AIBS 2009-11 grant proposals data with the lowest proportion of selected applicants, and the highest false negative rate of $0.27$ is for the FWF (1999-04) data set with the highest proportion of selected applicants ($53\%$).

These results highlight that most differences in false positive and false negative rates can be ascribed to the difference in the proportion of selected applicants. Although the false positive rates differ by 58 percentage points across the example data sets (37 percentage points excluding the lower IRR estimates of NSF \& COSPUP and AIBS), comparing the grant proposal selection procedures at an equal proportion of selected applicants (across the $5\% - 53\%$ range of selected applicants) reduces the maximum difference to 30 percentage points (15 percentage points excluding the lower IRR estimates of NSF \& COSPUP and AIBS; both at $5\%$ of selected applicants).

\begin{figure}[h!]
    \centering
    \includegraphics[width=0.6\textwidth]{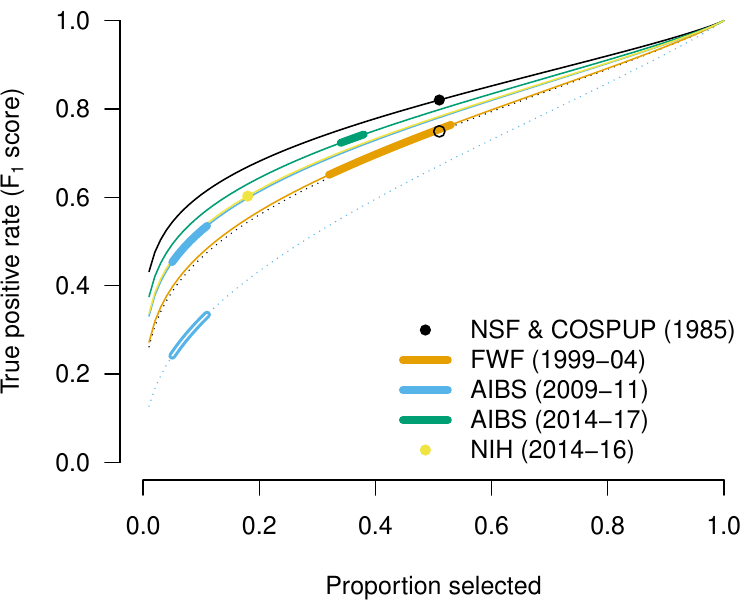}
    \caption{Estimates of the true positive rate (equivalent to $\text{F}_1$ score) for grant peer review procedures based on quantile approximation for published estimates of IRR and number of raters $J$ of individual studies (different colors). Thin full lines visualize the estimated false positive rate across the range of possible proportions of selected applicants, and thick full lines (and points) highlight the ranges (and values) of the actual proportion of selected applicants. For NSF \& COSPUP and AIBS, the lower $\text{IRR}_J$ estimate, resulting in a higher false positive rate, is shown as dotted lines / empty circles, whereas the higher $\text{IRR}_J$ estimate is shown as full lines / full circles.}
    \label{fig:example_F1}
\end{figure}

We can combine both types of error into an $\text{F}_1$ score, which in the case of selection procedures of the best candidates also corresponds to the true positive rate. Figure~\ref{fig:example_F1} visualizes the computed $\text{F}_1$ score based on the quantile approximation for each grant peer-review procedure in the same format as the previous figure. The NSF \& COSPUP (1985) grant proposals data result in the highest $\text{F}_1 = 0.82$ at the second highest proportion of observed candidates. The AIBS 2009-11 grant proposals data results in the lowest $\text{F}_1 = 0.24$ at the lowest proportion of selected applicants. Again, the large majority of differences between selection procedures are based on the proportion of selected candidates, however, the minimum difference between the best and worse performing selection procedure according to $\text{F}_1$ score remains $0.14$ ($0.09$ excluding the lower IRR estimates of NSF \& COSPUP and AIBS; both at $53\%$ of selected applicants).

\subsection{In Detail Assessment of NIH (2014-16)}

We can take a closer look at the NIH data set of \textcite{erosheva2021jrss} where the reported $\text{IRR}_1 = 0.34$ was accompanied by 95\% CI $[0.31, 0.37]$. The corresponding estimate of the false positive rate is $0.397$, and as in the simulation study, we can transform the 95\% CI of $\text{IRR}_1$ to obtain the 95\% CI interval $[0.380, 0.416]$ and the 95\% prediction interval $[0.331, 0.465]$ for the false positive rate. We can also compute the false negative rate estimate (i.e., type II error rate, Equation~\ref{eq:FNR}) of $0.087$, 95\% CI $[0.083, 0.091]$.

\begin{figure}[h!]
    \centering
    \includegraphics[width=1\textwidth]{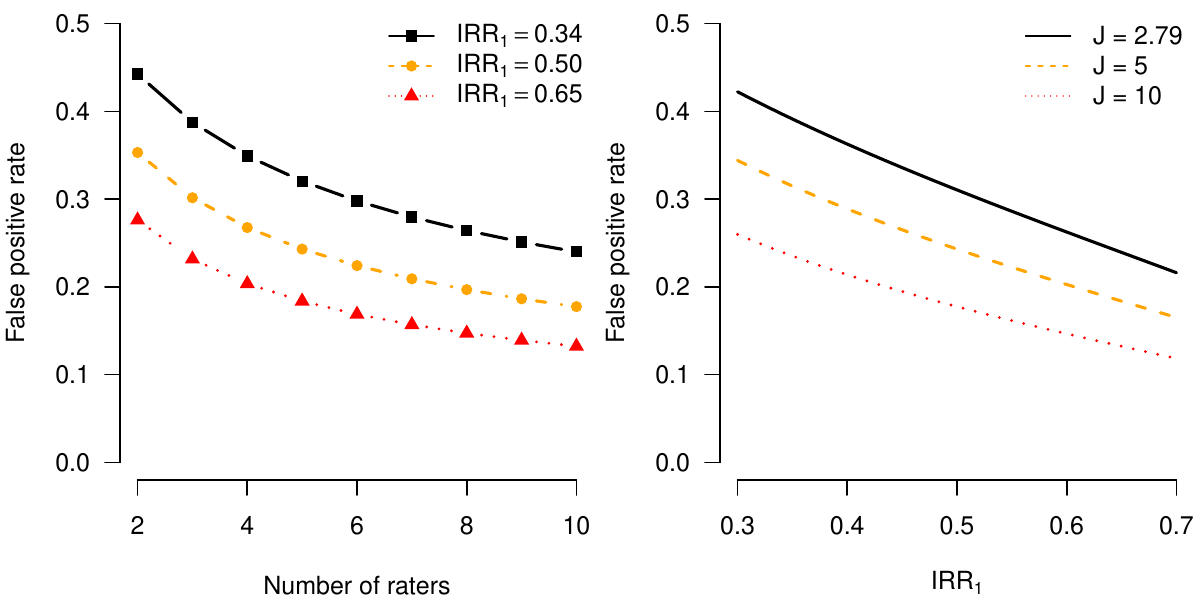}
    \caption{Change in false positive rate when increasing the number of raters for different levels of $\text{IRR}_1$ (left) or increasing the $\text{IRR}_1$ for different number of raters (right) for a selection procedure based on NIH (2014-16) grant peer review. The original trial selected $18$\% out of $2076$ applicants with $J = 2.79$ raters (black line in the right panel) and $\text{IRR}_1 = 0.34$ (black line in the left panel). Join increase of $\text{IRR}_1$ and the number of raters is depicted in orange and red.}
    \label{fig:example_design}
\end{figure}

We might redesign the selection procedure to achieve a lower false positive rate. This can be achieved by increasing the number of raters (left panel of Figure~\ref{fig:example_design}) or modifying the rating protocol to improve the inter-rater reliability $\text{IRR}_1$ (right panel of Figure~\ref{fig:example_design}).\footnote{The false positive and false negative rates are both based on the true classification probability and proportion of selected applicants---they cannot be optimized independently unless we consider changing the proportion of selected applicants as well.} Even though both changes lead to an increase in the overall inter-rater reliability, $\text{IRR}_J$, and consequently lower false positive rate, each change corresponds to a different modification of the peer review process (i.e., hiring more raters vs training the raters). 

The black line in Figure~\ref{fig:example_design} visualizes changes in the false positive rate when increasing the number of raters (left) or the $\text{IRR}_1$ (right). The added orange and red lines show an additional change in the false positive rate when altering both factors simultaneously. We see that for a selection procedure resulting in the selection of 18\% applicants, the false positive error rate remains relatively high even when increasing both the number of raters and $\text{IRR}_1$. Similar analysis can further incorporate the cost of each modification and weigh it against the benefit of decreasing the false positive rate of the selection procedure.

The calculations presented here are implemented in the \texttt{IRR2FPR} \texttt{R} package which can also be interactively run through the ShinyItemAnalysis application \parencite{martinkova2018shinyitemanalysis}.

\section{Discussion}
\label{sec:discussion}

We outlined an approach for evaluating ratings-based selection procedures in a binary classification framework. This approach allows researchers and stakeholders to assess the quality of selection procedures by linking inter-rater reliability to the actual selection decisions. As a result, the quality of selection procedures does not need to be evaluated only via a coefficient of inter-rater reliability but also via the desired false (true) positive/negative rates and their associated costs. These results can be used in more complex utility functions, featuring the costs of increasing the number of raters, modifying the rating guidelines, and changing the proportion of selected applicants.

The approach is based on a minimal measurement model, commonly used for assessing inter-rater reliability. We link the measurement model to binary decisions via a quantile approximation, assuming the goal of the selection procedure is selecting the best candidates. We showed how to compute the probability of correct classification and other binary classification metrics. We also showed how the binary classification relates to inter-rater reliability via the quantile approximation relying only on the number of ratings and proportion of selected applicants. The fact that the quality of the selection procedure evaluated via reliability is crucially dependent on the proportion of selected applicants echoes similar calls from past research on validity \parencite[][]{brogden1949testing, taylor1939relationship}.

The approach, however, only evaluates the selection procedures regarding reliability, assuming that the ratings are completely valid and the only limiting factor of the selection procedure is the measurement error \parencite{bornstein1991predictive}. That is, unfortunately, rarely the case. In cases with lower than perfect validity, the computed false positive rate can be used as a lower bound on the true false positive rate, which is further contaminated with low validity. Furthermore, one might argue against optimizing raters' reliability as the disagreement between raters might indicate that raters were selected on diverse bases and focused on different aspects of applicants \parencite[e.g.,][]{bailar1991reliability, hargens1991referee, kiesler1991confusion, lee2012kuhnian}.

We evaluated the quantile approximation in a simulation study and found little bias when the model assumptions were met. Breaking the dependency between the observed ratings and the latent abilities resulted in an overestimation of the true positive rate---which would result in the expected underestimation of the false-positive rate; however, a positive skew of the true abilities and dependency of the ratings on the raters led to different patterns of bias and increased RMSE. In general, the bias and RMSE decreased with an increasing number of raters, single inter-rater reliability, and the number of applicants. Importantly, the estimated classification probabilities and resulting binary classification metrics correspond to the expectation of the selection procedure. The actual classification probabilities for any given selection procedure might be far from the expectations, especially when considering only a small proportion of selected applicants---as the results are dependent on a few binary events that are inherently noisy.

We showed the methodology in an example comparing false positive rates,  false negative rates, and $\text{F}_1$ scores across multiple grant peer reviews. The results of our example should be interpreted with great caution. First, we based our results on averages across different fields that often vary in the inter-rater reliability, number of raters per proposal, and the number of selected applicants. Second, the grant agencies often base the funding decisions on a combination of the overall rating score and additional information (which is not included in our calculation). While acknowledging the limitations, the results still indicate a relatively high false positive rate and better controlled false negative rates due to the generally small portions of selected applicants. 

The outlined approach could be further expanded in multiple directions to accommodate different assumptions about the data-generating process. First, we used a minimal measurement model for assessing the inter-rater reliability of continuous ratings. However, each parameter of the measurement model might differ across groups of applicants \parencite[e.g.,][]{martinkova2018disparities, mutz2012heterogeneity, bartos2019testing, martinkova2023assessing}. Additional variance components might be used for modeling dependencies in the data \parencite[e.g., the effect of raters][]{martinkova2018disparities}. Different applicants might be rated a different number of times, the measured rankings might not be assumed to be normally distributed \parencite[e.g., see][for joint modeling of continuous and ordinal data]{pearce2022unified}, and the latent abilities might follow a different distribution as well. All these possibilities can be included in the measurement model and further propagated through the quantile approximation, either by using a mixture of distributions for the latent abilities based on the proportion of groups and group-specific parameters, specifying a different type of distributions and measurement error for the observed rankings, and different types of distributions for the latent abilities.

Second, we focused on classification probabilities and the resulting binary classification metrics themselves. Under the specified measurement model, most erroneous classifications happen close to the selection boundary. A further extension might consider weighting the classification probabilities by a utility function accounting for the degree of the error (i.e., mistakenly refusing an applicant who is right above the classification threshold is less costly than refusing an applicant much higher on the latent ability; see, e.g., \cite{cronbach1957psychological} for different utility models applied in the context of validity).

 Finally, in our real-data example, we focused on the context of the grant proposal's ratings; however, the explored connection between IRR and binary classification framework is relevant also in other areas. For example, in the context of educational measurement, the decisions about teacher hiring are based on an assessment of applicant submissions \parencite{goldhaber2021evidence}, the medical school admissions may be informed by ratings of personal and interpersonal qualities obtained in simulation-based assessment centers \parencite{ziv2008mor}, and the decisions about teacher promotion may be based on classroom observations \parencite{hill2012rater, casabianca2013effect}. In the context of psychological assessment and health-related measurements, the treatment assignment, as well as employment decisions, may be based on ratings from multiple raters or using multi-item instruments \parencite{rasova2012assessment}. Similarly, ratings from multiple raters inform the selection of journal articles \parencite{marsh1989peer}, and decisions in many other areas.

Despite its limitations, the presented approach provides a straightforward metric for evaluating the quality of selection procedures based on observable indicators. By considering this metric alongside other indicators, such as IRR, researchers, and stakeholders can gain a more comprehensive understanding of the quality of selection procedures.

\section*{Acknowledgement}

The study was funded by the Czech Science Foundation project "Theoretical Foundations of Computational Psychometrics" grant number 21-03658S, and by the project "Research of Excellence on Digital Technologies and Wellbeing CZ.02.01.01/00/22\_008/0004583" which is co-financed by the European Union. We thank Julius Pfadt and Elena Erosheva for their helpful comments and suggestions on an earlier version of the manuscript.

\section*{Supplementary material}

The analysis and simulation scripts as well as other electronic supplementary material are available at \url{https://osf.io/674fk/}. The \texttt{IRR2FPR} is available at \url{https://cran.r-project.org/package=IRR2FPR}.

\section*{Conflict of interests}

The authors declare that they have no conflict of interest.

\printbibliography
\newpage
\appendix
\section*{Appendix A: Further Simulation Results}
\label{app:simulation_results}

\begin{figure}[h!]
    \centering
    \includegraphics[width=\textwidth]{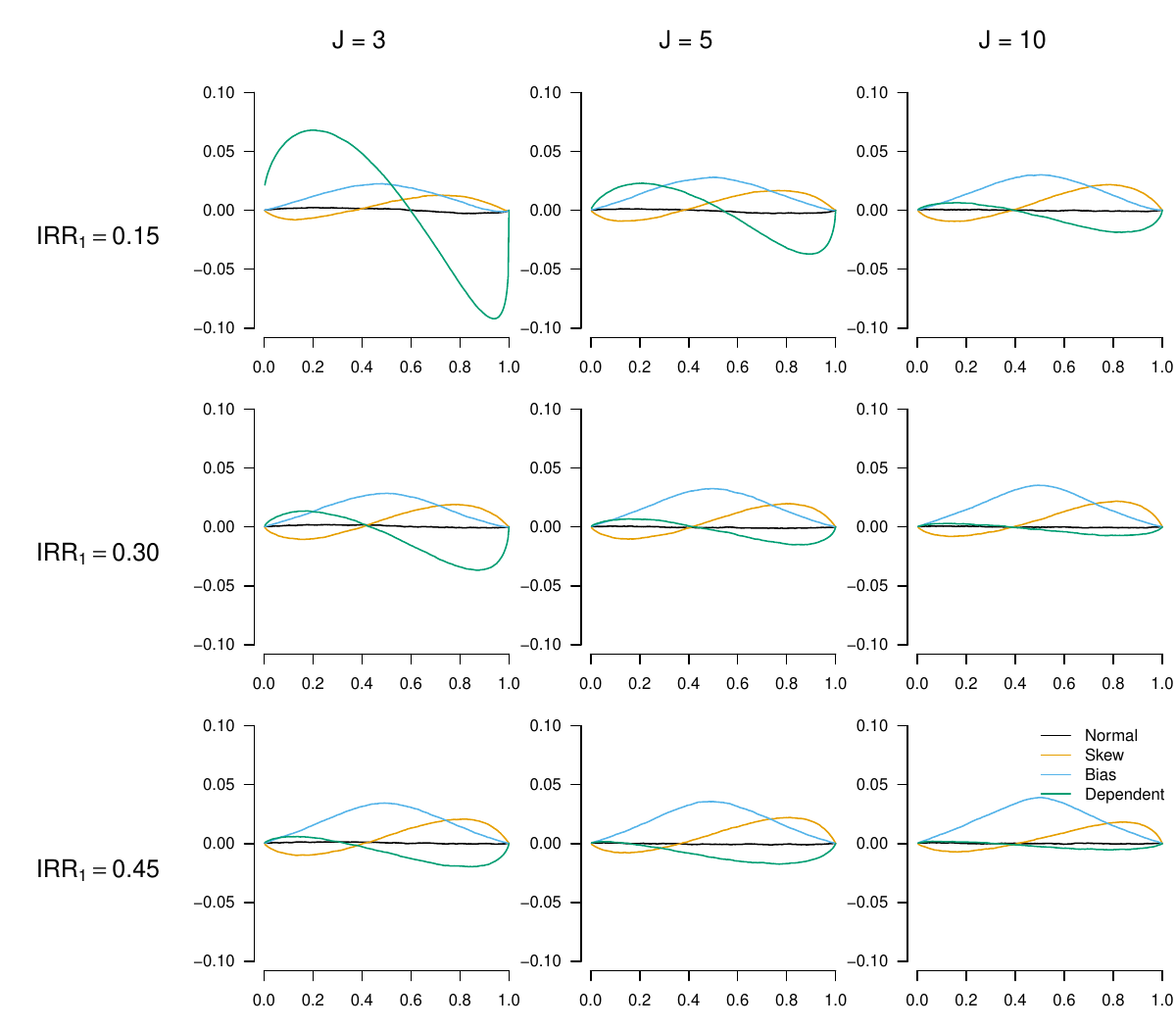}
    \caption{Bias of quantile approximated true positive classification probabilities for 300 applicants across all possible proportions of selected applicants ($x$-axis) with different data generating scenarios encoded with color. Results are presented for different inter-rater reliability coefficients (rows) and the number of ratings (columns).}
    \label{fig:sim_N=300}
\end{figure}

\begin{figure}[h!]
    \centering
    \includegraphics[width=\textwidth]{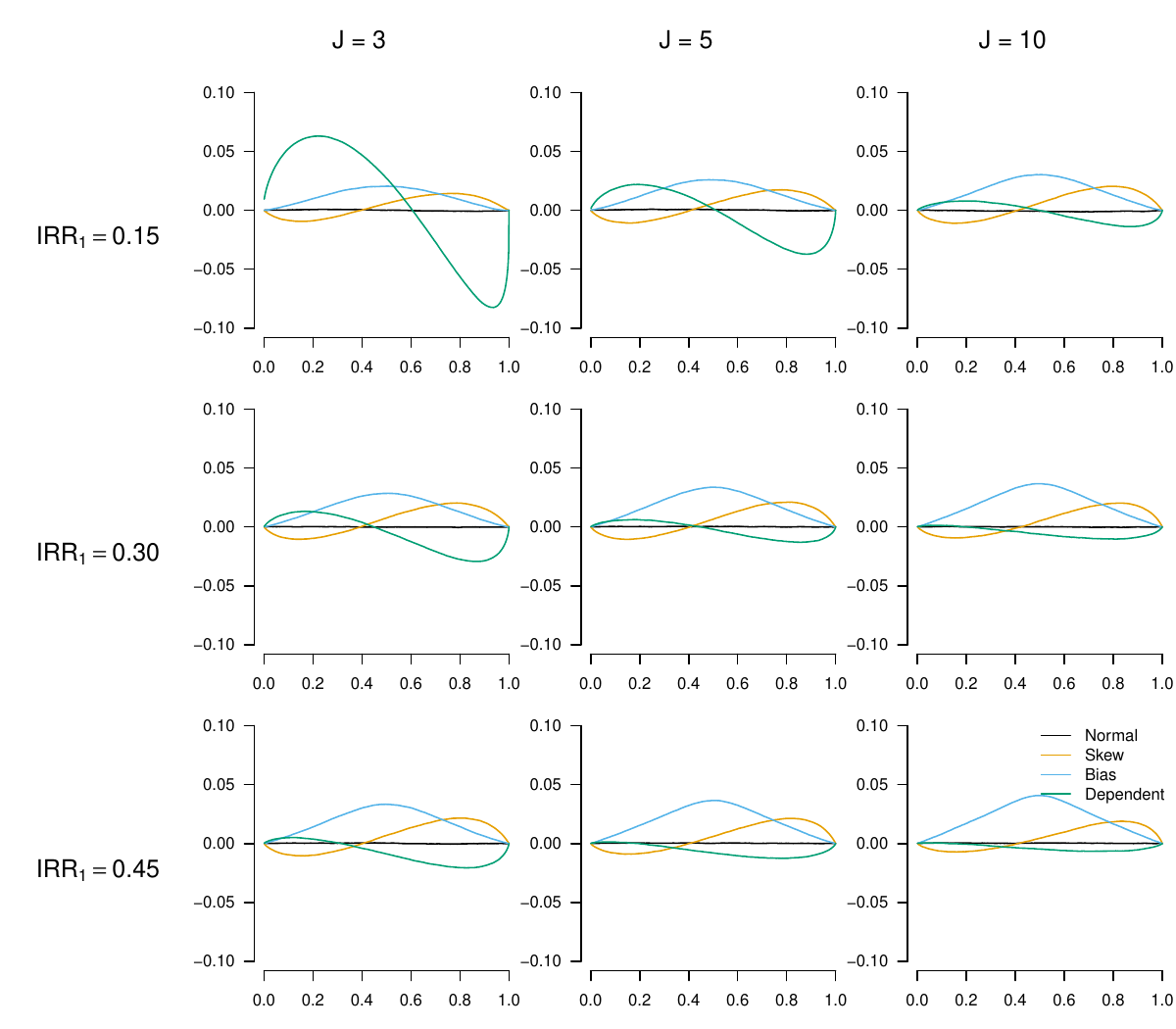}
    \caption{Bias of quantile approximated true positive classification probabilities for 1000 applicants across all possible proportions of selected applicants ($x$-axis) with different data generating scenarios encoded with color. Results are presented for different inter-rater reliability coefficients (rows) and the number of ratings (columns).}
    \label{fig:sim_N=1000}
\end{figure}

\begin{figure}[h!]
    \centering
    \includegraphics[width=\textwidth]{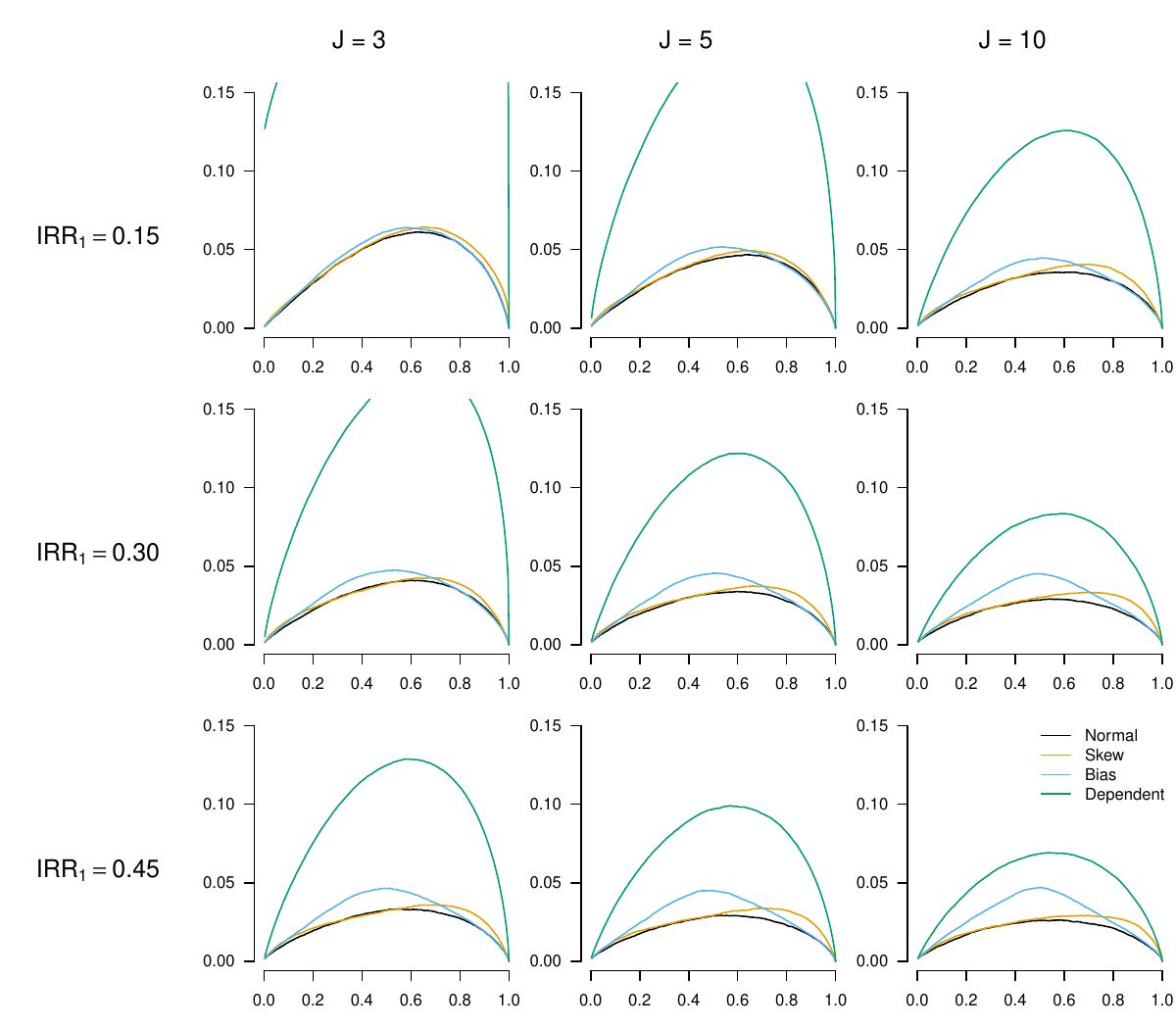}
    \caption{Root mean square error of quantile approximated true positive classification probabilities for 300 applicants across all possible proportions of selected applicants ($x$-axis)with different data generating scenarios encoded with color. Results are presented for different inter-rater reliability coefficients (rows) and the number of ratings (columns).}
    \label{fig:RMSE_N=300}
\end{figure}

\begin{figure}[h!]
    \centering
    \includegraphics[width=\textwidth]{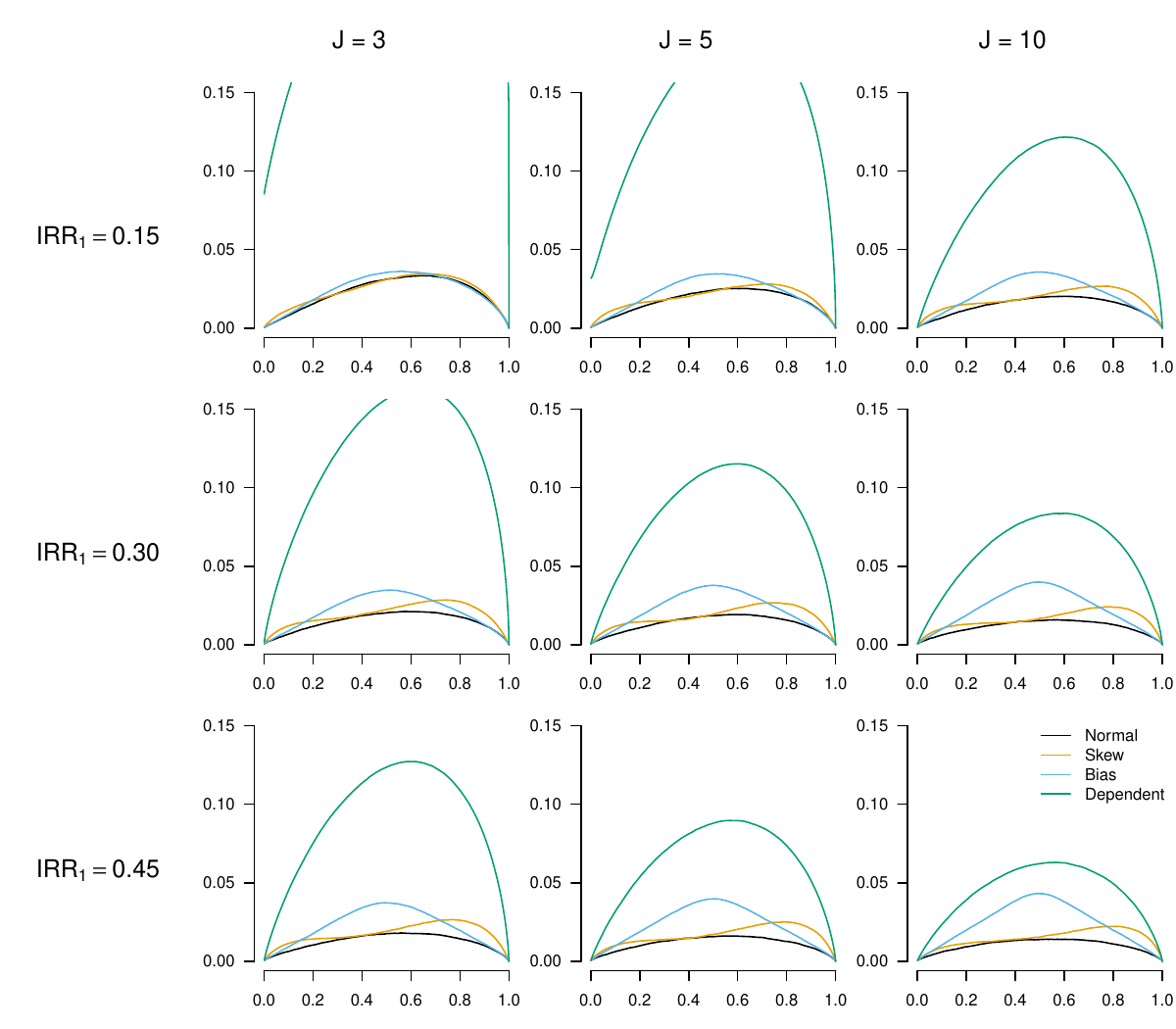}
    \caption{Root mean square error of quantile approximated true positive classification probabilities for 1000 applicants across all possible proportions of selected applicants ($x$-axis)with different data generating scenarios encoded with color. Results are presented for different inter-rater reliability coefficients (rows) and the number of ratings (columns).}
    \label{fig:RMSE_N=1000}
\end{figure}
\end{document}